\documentclass[aps,prd,twocolumn,showpacs,superscriptaddress,groupedaddress]{revtex4} 
\usepackage{mathrsfs}
\usepackage{epsfig}
\usepackage{graphicx}
\usepackage{mathtools}
\usepackage{color}
\usepackage{amsmath}

\newcommand{\tb}{\textbf}

\newcommand{\pr}{{\rm EDFs}}
\newcommand{\prs}{{\rm EDF}}
\newcommand{\prss}{{\rm EDF}}

\newcommand{\pp}{{\rm paper}}

\newcommand{\tj}[6]{ \begin{pmatrix}
   #1 & #2 & #3 \\
   #4 & #5 & #6 
  \end{pmatrix}}

    \newcommand{\mj}[6]{ \begin{bmatrix}
   #1 & #2 & #3 \\
   #4 & #5 & #6 
  \end{bmatrix}}
  
  \def\hp{ {\sc HEALPix}}

\begin{document}
\title{Detecting Electron Density Fluctuations from Cosmic Microwave Background Polarization using a Bispectrum Approach}

\author{Chang Feng\footnote{changf@illinois.edu}}
\affiliation{Department of Physics, University of Illinois at Urbana-Champaign, 1110 W Green St, Urbana, IL, 61801, USA}

\author{Gilbert Holder}
\affiliation{Department of Physics, University of Illinois at Urbana-Champaign, 1110 W Green St, Urbana, IL, 61801, USA}
\affiliation{Department of Astronomy, University of Illinois at Urbana-Champaign, 1002 W Green St, Urbana, IL, 61801, USA}
\affiliation{Canadian Institute for Advanced Research, Toronto, ON, Canada}

\begin{abstract}
Recent progress in high sensitivity Cosmic Microwave Background (CMB) polarization experiments opens up a window on 
large scale structure (LSS),  as CMB polarization fluctuations on small angular scales can arise from 
a combination of LSS and ionization fluctuations in the late universe. 
Gravitational lensing effects can be extracted from CMB datasets with quadratic estimators but reconstructions of electron density fluctuations (EDFs) with quadratic estimators are found to be significantly biased by the much larger
lensing effects in the secondary CMB fluctuations. In this \pp\ we establish a bispectrum formalism using tracers of
LSS to extract the subdominant \pr\ from CMB polarization data.  We find that this bispectrum can effectively reconstruct angular band-powers of cross correlation 
between \pr\ and LSS tracers. Next generation CMB polarization experiments in conjunction with galaxy surveys and cosmic infrared background experiments can detect signatures of \pr\ with high significance.
\end{abstract}

\maketitle

\section{Introduction}
Secondary fluctuations of the comic microwave background (CMB) arise from gravitational perturbations and Compton
scattering that occur after the epoch of recombination, as the relic photons traverse a universe that has spatial
and temporal variations in the gravitational potential and electron density.
These secondary fluctuations are rich in information about structure formation in the universe, and are 
becoming invaluable probes for studying large scale structure (LSS) and physical processes during the epoch of reionization  
and in the recent universe.

Quadratic estimators have been devised to map the matter distribution through CMB gravitational lensing effects~\cite{2001PhRvD..64h3005H,2003PhRvD..67l3507K}, and this 
has been detected at very high significance from both CMB temperature ($T$) and polarization ($E$ and $B$)~\cite{2007PhRvD..76d3510S,2008PhRvD..78d3520H,2012PhRvD..86f3519F,2011PhRvL.107b1301D,2013PhRvL.111n1301H,2014PhRvL.112m1302A,2014PhRvL.113b1301A}. 
Quadratic estimators for electron density fluctuations (EDFs) have been applied to CMB data sets and no detection has been made to date, indicating that the \prss\ signal is subdominant to CMB lensing.
For example, the $TB$ quadratic estimator was applied to Wilkinson Microwave Anisotropy Probe 7-year data and no signal of \pr\ was found~\cite{2013PhRvD..87d7303G}. This analysis was recently updated with the Planck 2015 temperature data~\cite{2017arXiv171100058N}; the constraint on \pr\ was greatly improved but no signal was detected. 

Using numerical simulations, previous works have investigated reconstruction of the \pr\, 
with a $EB$ flat-sky estimator~\cite{2011arXiv1106.4313S} and a full sky $EB$ estimator~\cite{cora1}. 
The reconstruction formalism of \pr\ resembles CMB lensing but both studies~\cite{2011arXiv1106.4313S,cora1} found significant biases in reconstructed power spectra. The secondary CMB fluctuations induced by \prs\ signal is expected to be
orders of magnitude smaller than those induced by CMB lensing, and the first-order approximation for the \prs\ reconstruction 
is not sufficient. Higher order corrections to both \pr\ and CMB lensing are not negligible in the four-point correlation 
functions that arise when calculating these power spectra.

In this \pp, we develop a bispectrum technique using tracers of LSS to both extract \prss\ signal and delens 
the CMB simultaneously. This bispectrum technique is a generic formalism that can be applied to 
other subdominant secondary effects.

\section{Bispectra generated by gravitational potential and electron density fluctuations}

The bispectrum can be generally defined as
\begin{equation}
\hat b_{\ell_1\ell_2\ell_3}=\displaystyle\sum_{m_1m_2m_3}\tj{\ell_1}{\ell_2}{\ell_3}{m_1}{m_2}{m_3}\langle a_{\ell_1m_1}a_{\ell_2m_2}a_{\ell_3m_3}\rangle\label{def}
\end{equation}
for any modes $a_{\ell m}$ where $a_{\ell m}$s are spherical harmonic coefficients and $(...)$ is the 3-$j$ Wigner symbol. The CMB in the sky direction ${\textbf n}$ with both the gravitational lensing ($\phi$) effect and spatially varying optical depth ($\tau=\tau_0+\delta\tau$) can be described as
\begin{equation}
X({\bf n})=\tilde X({\bf n}+\nabla\phi({\bf n}))e^{-\tau({\bf n})}
\end{equation}
which can be expanded to leading order in both $\nabla \phi$ and $\tau$. Here $X$ can be observed CMB temperature ($T$), and Stokes parameters ($Q$ and $U$) for the CMB polarization, while the unlensed and unscreened CMB field is $\tilde X$. In the rest of the paper, we neglect the constant part $e^{-\tau_0}$ and refer $\delta\tau$ to $\tau$ for simplicity. Also, we use the electric- ($E$) and curl-like ($B$) modes, which are converted from the Stokes parameters, for the analysis. In addition to primordial and gravitational lensing $B$ modes, CMB polarization fluctuations can also be generated by screening effect and Thomson scattering of the local temperature quadrupole. Following~\cite{2011arXiv1106.4313S,2013PhRvD..87d7303G,2017arXiv171100058N}, we focus on a reconstruction of spatially varying optical depth with $B$ modes by gravitational lensing and screening effect and defer a comprehensive discussion with all possible $B$-mode sources in future work. Symbolically, the bispectrum involving observed CMB fields ($X$ and $Y$) and a tracer map $\Psi$ can be split into two components as
\begin{equation}
\langle XY\Psi\rangle\propto C_{\ell}(X,Y)C_{\ell'}^{\phi\Psi} + C_{\ell}(X,Y)C_{\ell'}^{\tau\Psi}.
\end{equation}

Therefore, the two secondary effects would non-negligibly contribute to the overall bispectrum. Using the bispectrum definition in Eq. (\ref{def}) and specializing to polarization $E$ and $B$ modes, we express the two component-separated reduced bispectra as

\begin{equation}
b^{(\phi)}_{\ell_1\ell_2\ell_3}(E,B,\Psi)=\tilde C_{\ell_1}^{EE}C_{\ell_3}^{\phi\Psi}\mj{\ell_2}{\ell_3}{\ell_1}{\pm 2}{0}{\mp 2}\Pi_{\ell_1\ell_2\ell_3}\xi_{\ell_1\ell_2\ell_3},\label{3p-phi}
\end{equation}
and
\begin{equation}
b^{(\tau)}_{\ell_1\ell_2\ell_3}(E,B,\Psi)=\tilde C^{EE}_{\ell_1}C_{\ell_3}^{\tau\Psi}\mj{\ell_2}{\ell_3}{\ell_1}{\pm 2}{0}{\mp 2}\Pi_{\ell_1\ell_2\ell_3}.\label{3p-tau}
\end{equation}
Here $\xi_{\ell_1\ell_2\ell_3}=[\ell_1(\ell_1+1)+\ell_3(\ell_3+1)-\ell_2(\ell_2+1)]/2$ and $\Pi_{\ell_1\ell_2\ell_3}=\sqrt{(2\ell_1+1)(2\ell_2+1)(2\ell_3+1)/(4\pi)}$. For spin-2 fields, we also define a hybrid 3-$j$ Wigner symbol as
\begin{equation}
\mj{\ell}{L}{\ell'}{\pm2}{0}{\mp2}=\frac{1}{2i}\Big[\tj{\ell}{L}{\ell'}{2}{0}{-2}-\tj{\ell}{L}{\ell'}{-2}{0}{2}\Big].
\end{equation}
From lengthy derivations, we find that the general bispectrum estimators involving two arbitrary CMB modes have very similar mathematical structures as Eqs. (\ref{3p-phi}, \ref{3p-tau}).

The CMB related power spectra can be conveniently computed by CAMB~\cite{2000ApJ...538..473L}, and we 
use the halo model formalism~\cite{2002PhR...372....1C} to calculate all the theoretical auto- and cross-power spectra among $\phi$, $\tau$ and $\Psi$, forming six power spectra $C_{\ell}^{\phi\phi}$, $C_{\ell}^{\tau\tau}$, $C_{\ell}^{\Psi\Psi}$, $C_{\ell}^{\phi\tau}$, $C_{\ell}^{\phi\Psi}$ and $C_{\ell}^{\tau\Psi}$. For simplicity, we chose
the CAMB's reionization model~\cite{2008PhRvD..78b3002L}, and self-consistently derived the reionization history according to the latest Planck constraint $\tau=0.058\pm0.012$~\cite{2016A&A...596A.108P}. To leading order, the \pr\ are linearly proportional to density contrasts of both matter and ionizing fields. The statistical properties of the ionizing field are determined by a bubble model~\cite{2004ApJ...613....1F,2006ApJ...643..585W} in which we assume that the bubble size satisfies a simple logarithmic distribution and the characteristic variance of the bubble size is set to unity. The redshift-dependent bubble size is self-consistently solved when the ionization fraction is given. In practice, using LSS tracers, we find that the signal is dominated by density fluctuations rather than the patchy signal from reionization. Also, it is assumed that helium is singly ionized along with hydrogen while the double ionization of helium is neglected. More detailed discussions of the reionization model can be found in~\cite{2017ApJ...846...21F}. In this \pp\ we only focus on hydrogen reionization and the approach discussed in this work can be easily applied to helium reionization as well. Second reionization of helium provides a particularly interesting application of this formalism, which we defer to future work. 

We assume a survey like Large Synoptic Survey Telescope (LSST)~\cite{2009arXiv0912.0201L} and adopt the redshift distribution from Ref.~\cite{2017arXiv171009465S} for the galaxy density contrast modeling. In addition, the broad redshift distribution of the cosmic infrared background (CIB) traces a substantial portion of the large scale structure in the late universe so it is also a sensitive tracer field $\Psi$. We use the Planck CIB model~\cite{2014A&A...571A..30P} and calculate all the aforementioned angular power spectra at 857 GHz (350 $\mu${\rm m}) as we do for the galaxy surveys, assuming a full-sky CIB experiment with negligible instrumental noise.

Internally, a direct map of $\phi$ from CMB lensing reconstruction can be chosen as a tracer field $\Psi$ but the signal-to-noise ratio per pixel is not high enough to trace the \prss\ signal even with CMB-Stage 4 (CMB-S4)~\cite{2016arXiv161002743A} datasets.
More importantly, the lensing reconstruction of the observed CMB maps might contain a small \prss\ signal~\cite{2011arXiv1106.4313S}, which can contaminate the reconstruction of the $\tau$-type bispectrum in Eq. (\ref{3p-tau}), so we use the external tracers -- galaxy number count ($g$) and the CIB ($\Theta$).

We perform the Cholesky decomposition of the covariance matrix among $\phi$, $\tau$ and $\Psi$ to generate correlated Gaussian simulations. We numerically validate that all the six power spectra calculated from the correlated simulations can exactly recover the input ones. The noise realizations are also generated for different tracers and CMB polarization maps.

Using all the theoretical power spectra and full-sky simulations, we first run \textit{Taylens}~\cite{2013JCAP...09..001N} to make lensed CMB polarization simulations and then perform $X({\bf n})=\tilde X({\bf n})e^{-\tau({\bf n})}$ to encode a \prss\ field for each lensed realization. Moreover, we create unlensed, lensed ($\phi$-only), $\tau$-only and full CMB polarization simulations that contain both $\phi$ and $\tau$ fields. From the $\phi$- and $\tau$-only simulations, it is seen that the excess power generated by \pr\ is about three orders of magnitude smaller than that of gravitational lensing, verifying that the signal of \pr\ can not be extracted at the power spectrum level and estimators with higher order correlation functions, such as the bispectrum, are required. 

\begin{figure*}
\includegraphics[width=8cm, height=6cm]{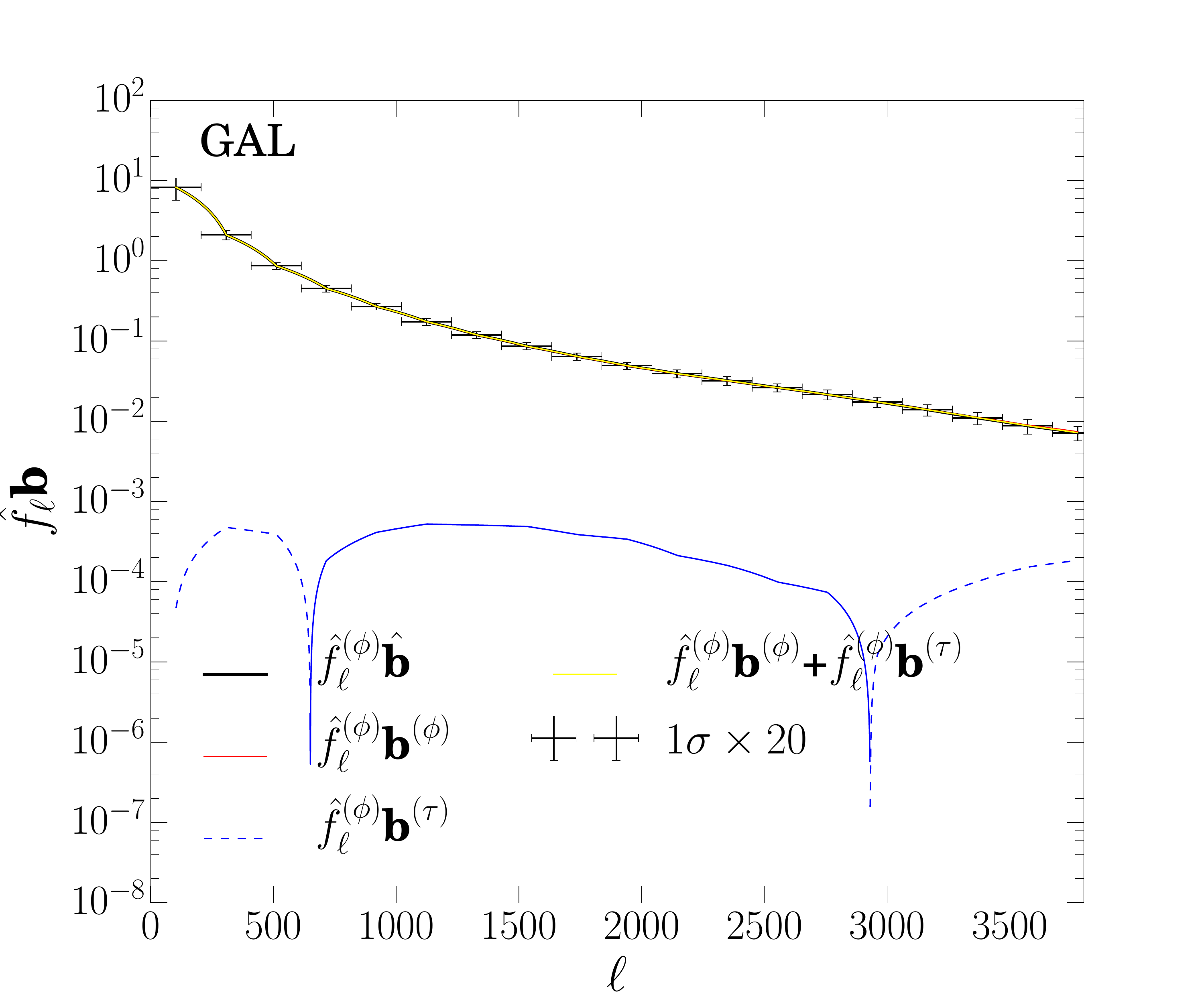}
\includegraphics[width=8cm, height=6cm]{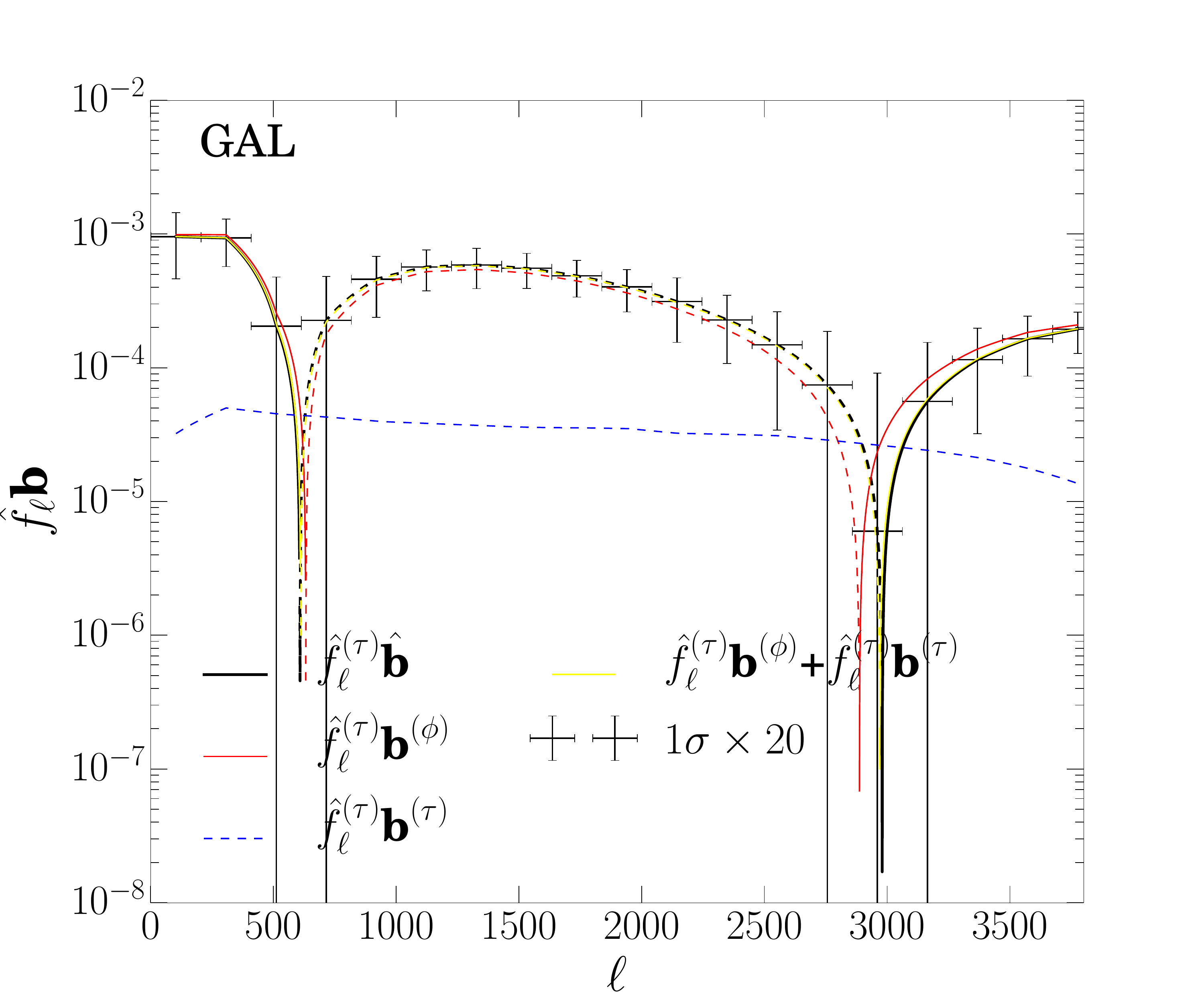}
\includegraphics[width=8cm, height=6cm]{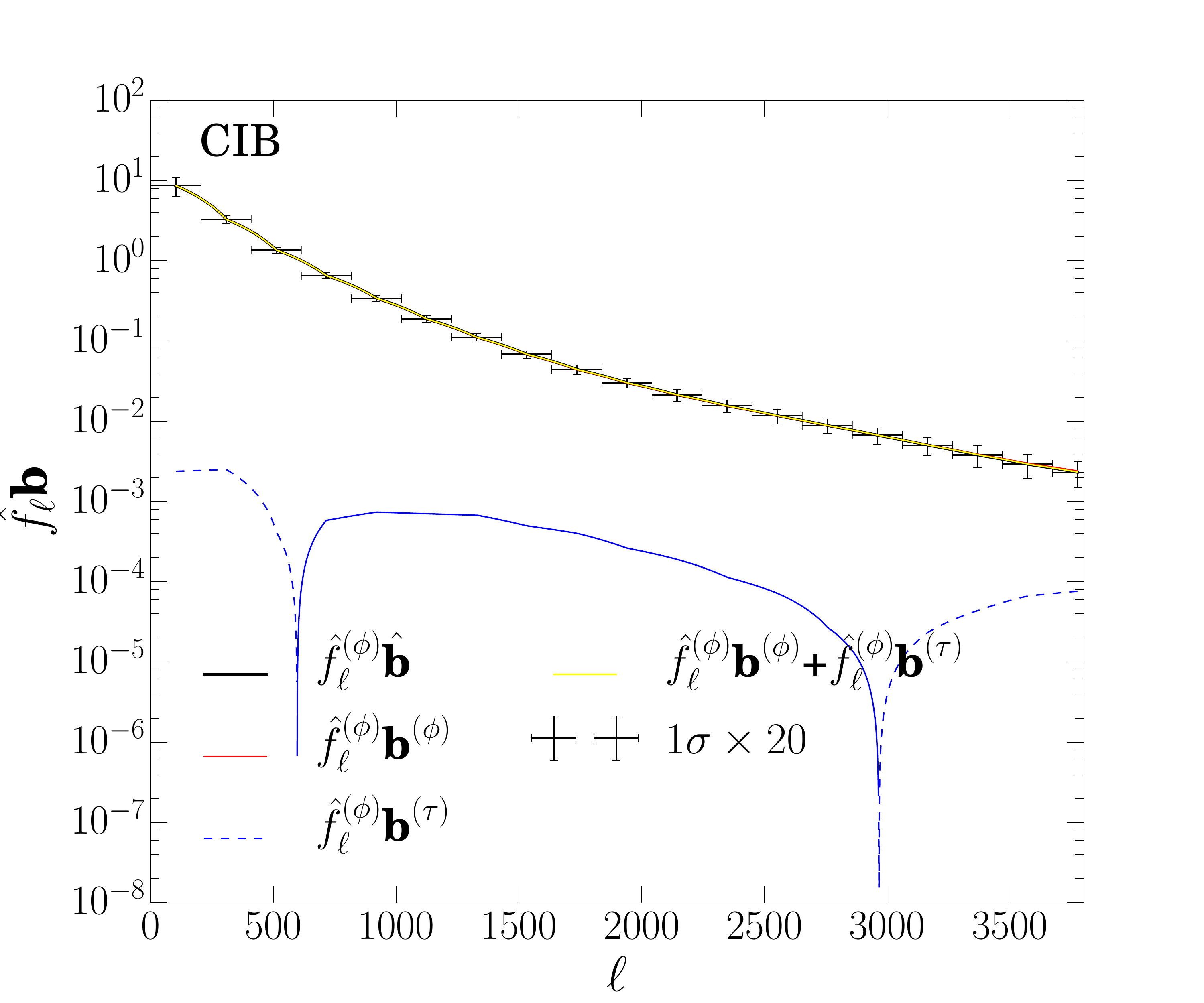}
\includegraphics[width=8cm, height=6cm]{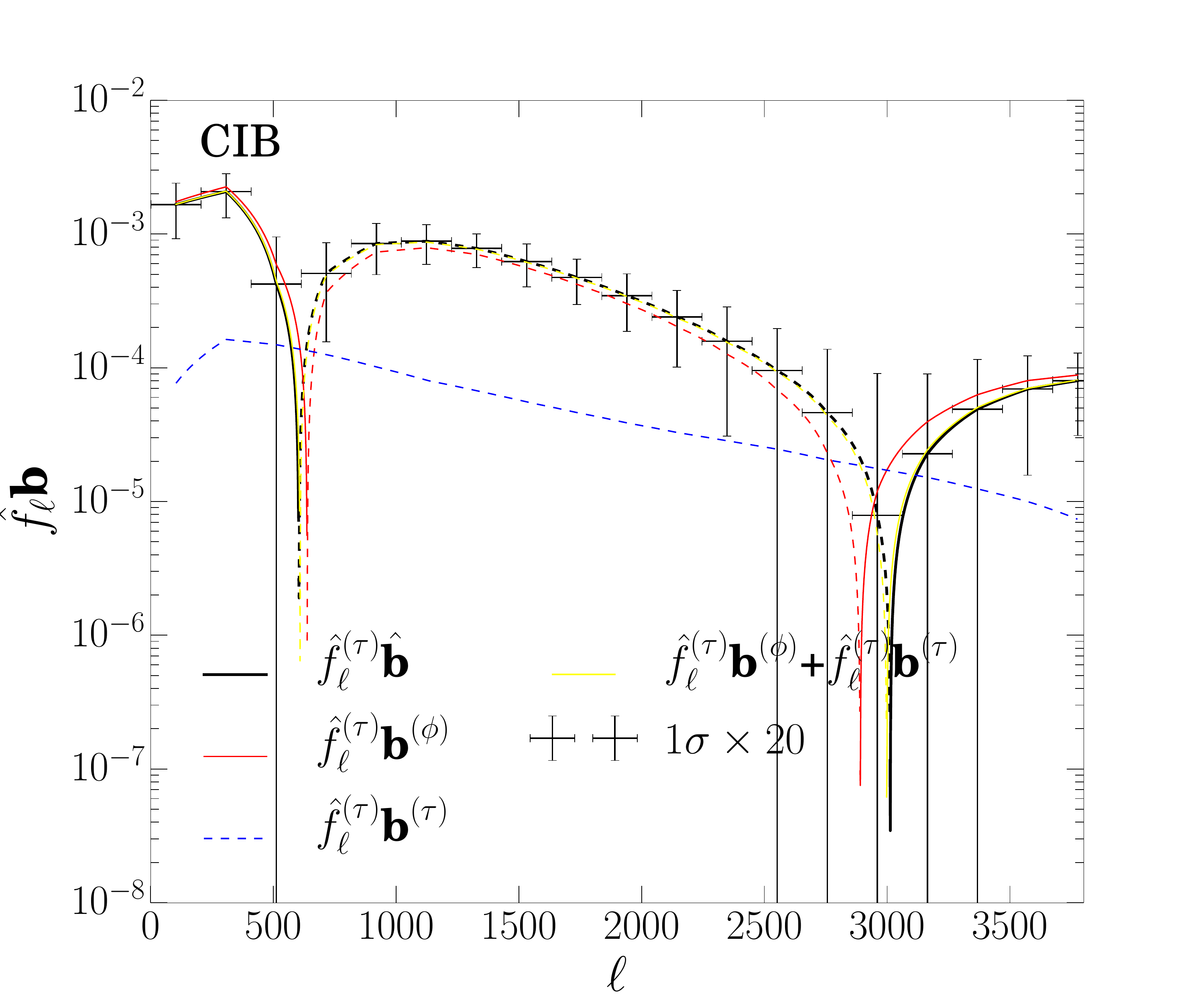}
\caption{Raw bispectra from CMB lensing and electron density fluctuations. The top and bottom panels correspond to raw bispectra with galaxy and CIB tracers, respectively. (Left):$\phi$-channel bispectrum components (first row of Eq. (\ref{ML})); (Right): $\tau$-channel bispectrum components (seconds row of Eq. (\ref{ML})). The dashed portions are negative. The simulations are done at \textit{Healpix} resolution $N_{\rm side}=2048$ and $\ell_{\rm max}=4096$ with $f_{\rm sky}$ = 1, $\Delta_P$ = $\sqrt{2}\,\mu{\rm K}\mbox{-}{\rm arcmin}$ and $\theta_{\rm FWHM}$ = 1$^{\prime}$. In the left panel, $\phi$-type bispectrum is about three orders of magnitude higher than the $\tau$-type, but in the right panel, the $\tau$-type bispectrum is substantially amplified, giving rise to a detectable $\langle\tau \Psi\rangle$ signal.}\label{raw}
\end{figure*}

We construct maximum likelihood estimators
\begin{equation}
\begin{bmatrix}
\hat f^{(\phi)}_{\ell}\hat {\tb{b}} \\
\hat f^{(\tau)}_{\ell}\hat {\tb{b}}
\end{bmatrix}
=\begin{bmatrix}
\hat f^{(\phi)}_{\ell}{\tb{b}}^{(\phi)} & \hat f^{(\phi)}_{\ell}\tb{ b}^{(\tau)} \\
\hat f^{(\tau)}_{\ell}{\tb{b}}^{(\phi)} & \hat f^{(\tau)}_{\ell}\tb{ b}^{(\tau)}
\end{bmatrix}
\begin{bmatrix}
A^{\phi\Psi}_{\ell} \\
A^{\tau\Psi}_{\ell}
\end{bmatrix}\label{ML}
\end{equation}
to reconstruct both $\langle\phi\Psi\rangle$ and $\langle\tau\Psi\rangle$ power spectra simultaneously. Here we define an operator
\begin{equation}
\hat f^{(i)}_{\ell}=\displaystyle\sum_{\ell_1\ell_2}\frac{b^{(i)}_{\ell\ell_1\ell_2}}{C_{\ell}C_{\ell_1}C_{\ell_2}}\label{op}
\end{equation}which acts on a bispectrum. The amplitude $A^{c\Psi}_{\ell}$ is the ratio between the reconstructed and the input band-powers $C^{c\Psi}_{\ell}$ and $c=\phi$ or $\tau$. The bispectrum $\hat{\textbf b}$ can be calculated from either real or mock polarization data with both $\phi$ and $\tau$ signals. The simulations are used to derive the right-hand side which consists of $\tb{b}^{(\phi)}$ and $\tb{b}^{(\tau)}$ that are constructed by $\langle E_{\ell m}B^{(\phi)}_{\ell'm'}\Psi_{\ell''m''}\rangle$ and $\langle E_{\ell m}B^{(\tau)}_{\ell'm'}\Psi_{\ell''m''}\rangle$, respectively. Here $B^{(\phi)}$ and $B^{(\tau)}$ are CMB B modes generated from $\phi$- and $\tau$-only simulations.
Based on this method, we optimally decouple $\phi$- and $\tau$-type bispectra, equivalently, extracting the signal of \pr\ while delensing the CMB polarization data.

Ideally, one can subtract the lensing-induced bispectrum (Eq.(\ref{3p-phi})) and obtain the power spectrum from the delensed bispectrum that is only generated by the \pr. We will show in the next sections that the lensing-induced bispectrum is significantly larger than the $\tau$-induced one, so a tiny uncertainty on the lensing bispectrum would result in a significant contamination on the signal of \pr. This indicates that a nearly perfect delensing of the CMB maps will be required. In principle, one can first delens the CMB polarization data and then use the delensed one to construct a bispectrum estimator which eliminates the $\phi$-type in Eq. (\ref{ML}) but the delensing efficiency is not equal to 100\% at all angular scales so such a two-step approach still leaves some residual of the lensing signal in the CMB polarization data. This can easily contaminate the much fainter signal of \pr\ which is about three orders of magnitude weaker than CMB lensing in terms of CMB angular power. 
With the approach described by Eq. (\ref{ML}), we do not need extra efforts to delens the data, but to extract both lensing and \prs\ signals simultaneously. 

The summation in Eq. (\ref{op}) would be very time-consuming if $\ell_{\rm max}$ is very large. We adopt an efficient algorithm to compute the operator $\hat f_{\ell}{\textbf b}$ which can be mathematically factored into three weighted maps. Specifically, the efficient estimator is re-written by real-space map products as 

\begin{eqnarray}
\hat f^{(c)}_{\ell}{\tb{b}}^{(c)}\delta_{\ell\ell'}\delta_{mm'}&=&\displaystyle\sum_{i}\langle\mathcal{F}^{-1}[{}_{\pm2}X^{(c,i)}({\bf n}){}_{\mp2}Y^{(c,i)}({\bf n})]_{\ell m}\nonumber\\
&\times&Z_{\ell' m'}^{(c,i)}\rangle.
\end{eqnarray}
This efficient procedure can be applied to all the terms in Eq. (\ref{ML}) and we use $\phi$-only and $\tau$-only simulations to generate two types of bispectra ${\tb{b}}^{(c)}$. In this equation, $\mathcal{F}^{-1}$ refers to spin-0 inverse spherical harmonic transformation, $c$ = $\phi$ or $\tau$, $i$ is the index listed in Tables \ref{phiw} and \ref{tauw}, 
${}_{\pm2}X^{(c,i)}({\bf n})$ = $\displaystyle\sum_{\ell m}\alpha_{\ell}^{(c,i)}$$E_{\ell m}\,{}_{\pm2}Y_{\ell m}$, $
{}_{\mp2}Y^{(c,i)}({\bf n})$ = $\displaystyle\sum_{\ell m}\beta_{\ell}^{(c,i)}$$B_{\ell m}\,{}_{\mp2}Y_{\ell m}$, and
${}_{0}Z^{(c,i)}({\bf n})$ = $\displaystyle\sum_{\ell m}\gamma_{\ell}^{(c,i)}$$\Psi_{\ell m}\,{}_{0}Y_{\ell m}$. All the filtering functions $\alpha$, $\beta$ and $\gamma$ are given in Tables \ref{phiw} and \ref{tauw} for $\phi$- and $\tau$-type bispectrum estimators. The CMB noise power spectra are included in the denominator of the filters in the Tables, and $N_{\ell}^{\Psi\Psi}$ denotes either the shot noise of the galaxy survey or instrumental noise for the CIB experiment.

\begin{figure*}
\includegraphics[width=8cm, height=6.7cm]{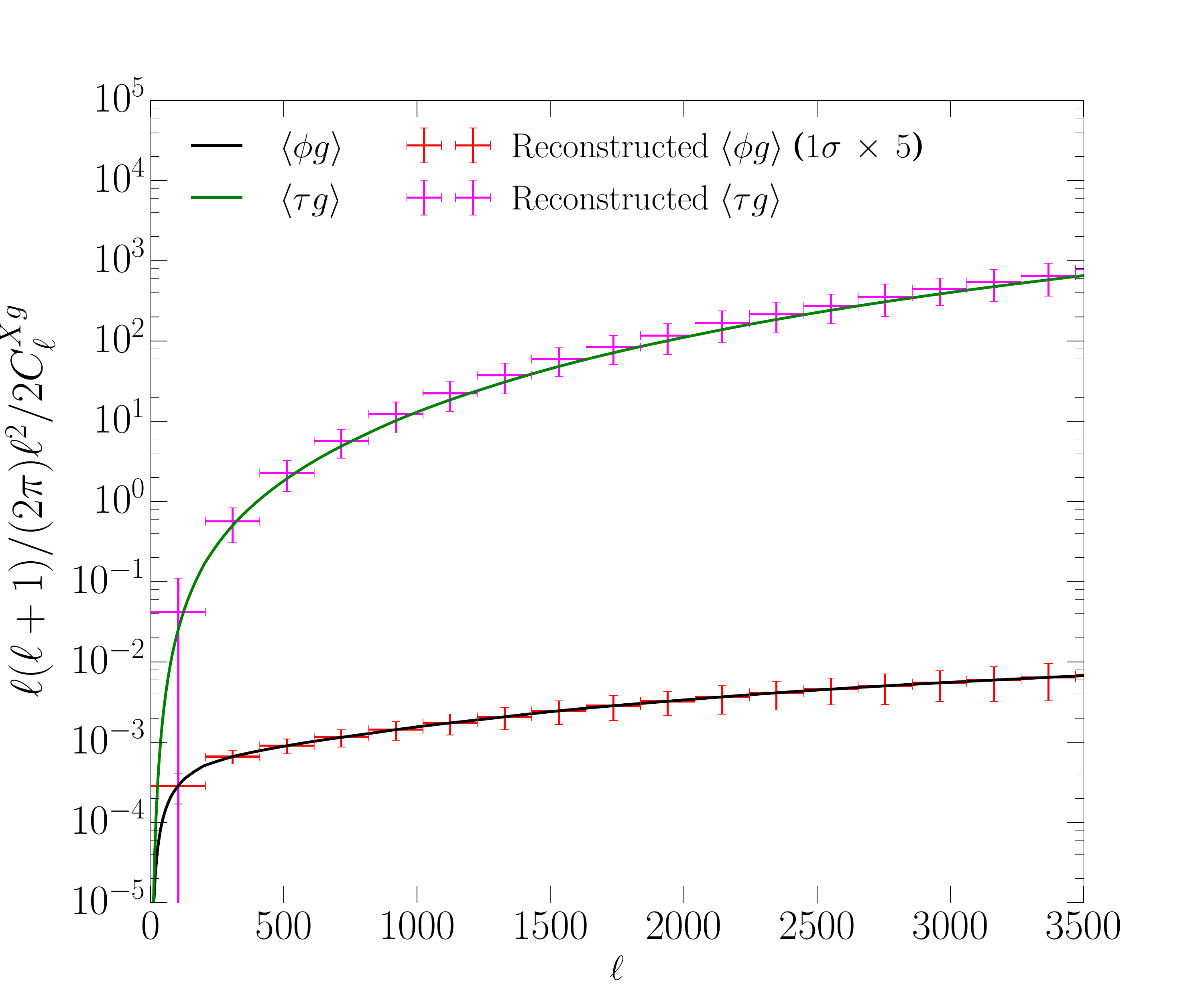}
\includegraphics[width=8cm, height=6.7cm]{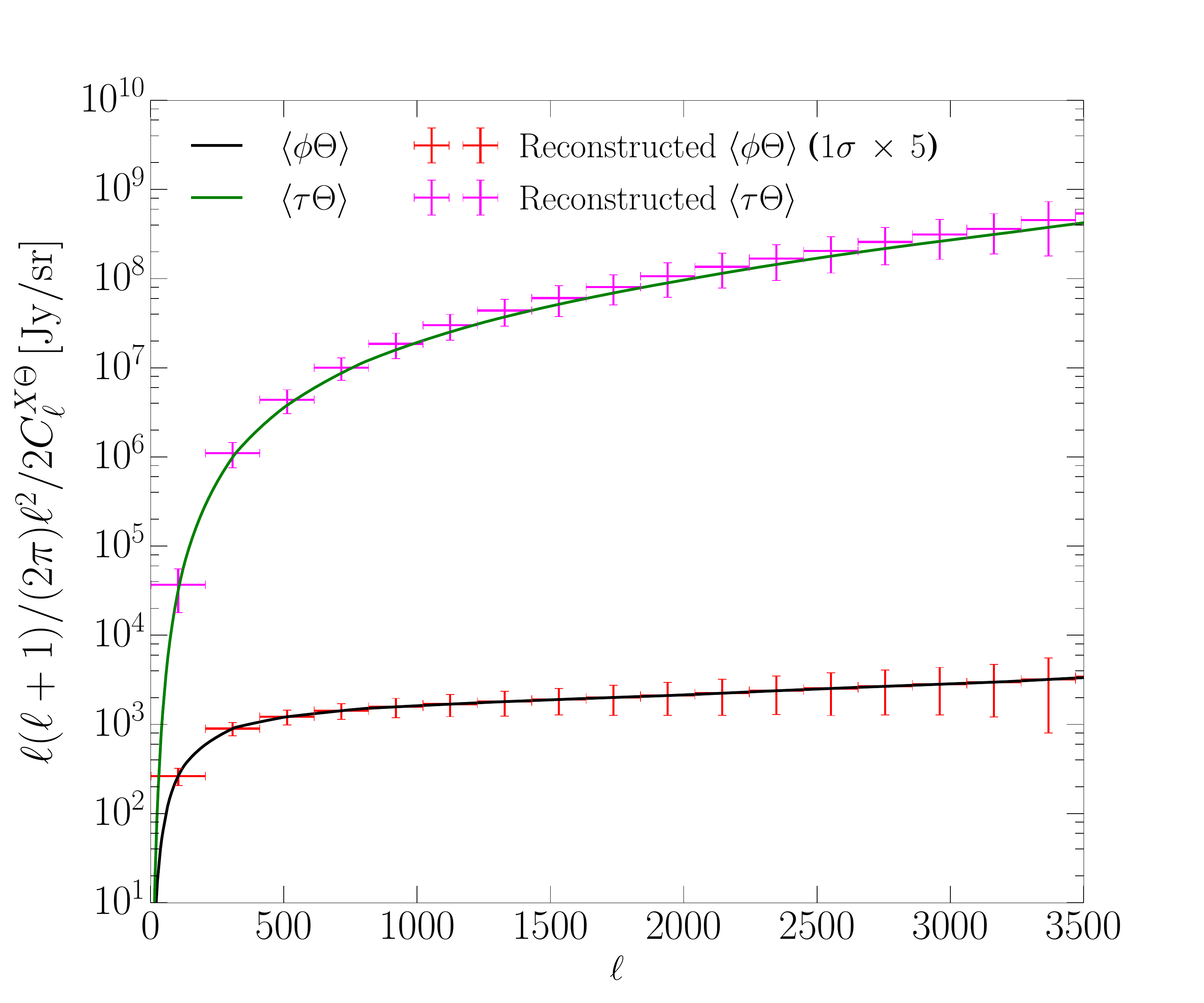}
\caption{The reconstructed $\langle\phi\Psi\rangle$ and $\langle\tau\Psi\rangle$ cross-power spectra from $\langle EB\Psi \rangle$ bispectrum estimator. The symbol $X$ in the $y$-axis denotes either $\phi$ or $\tau$. (Left) reconstructed band-powers $\langle\phi g\rangle$ and $\langle\tau g\rangle$;  (Right) reconstructed band-powers $\langle\phi\Theta\rangle$ and $\langle\tau\Theta\rangle$. The simulations are done at \textit{Healpix} resolution $N_{\rm side}=2048$ and $\ell_{\rm max}=4096$ with $f_{\rm sky}$ = 1, $\Delta_P$ = $\sqrt{2}\,\mu{\rm K}\mbox{-}{\rm arcmin}$ and $\theta_{\rm FWHM}$ = 1$^{\prime}$. A LSST-type galaxy survey and a full-sky CIB measured at 857 {\rm GHz} are assumed.}\label{recon}
\end{figure*}

From simulations we find that the temperature-related bispectrum estimators for the next generation CMB experiments and LSS surveys have much smaller signal-to-noise ratios than polarization-related estimator so we do not include the four pairs $TT\Psi$, $TE\Psi$, $TB\Psi$ and only focus on $EB\Psi$. We will discuss all the six estimators, as well as the minimum variance estimator, in future work.

\section{Results}

In Figure \ref{raw}, we show different components of the $EB\Psi$ bispectrum. The black lines and the simulated band powers in the left and right panels are the two terms on the left-hand side of Eq. (\ref{ML}), i.e., the raw bispectra filtered by the $\phi$- and $\tau$-type channels, respectively. The red and light blue lines in each panel are $\phi$- and $\tau$-type bispectra filtered by each channel, corresponding to each row of the right-band side of Eq. (\ref{ML}). It is clearly seen that the $\phi$-type bispectrum is about three orders of magnitude higher than the $\tau$-type in the left panel ($\phi$-channel), but in the right panel, the $\tau$-type bispectrum is substantially amplified in the $\tau$-channel and the $\phi$-type bispectrum is significantly suppressed. The sum of the $\phi$- and $\tau$-type bispectra in each panel is equal to the raw bispectrum, as the yellow lines show so the bispectrum estimators in Eq. (\ref{ML}) are fully validated as unbiased. The reconstructed band powers $C_{\ell}^{\phi \Psi}$ and $C_{\ell}^{\tau \Psi}$ are shown in Figure \ref{recon} for the two tracers -- galaxy number count and CIB, and no significant biases are seen at all the angular scales.

\begin{figure*}
\includegraphics[width=8cm, height=7.2cm]{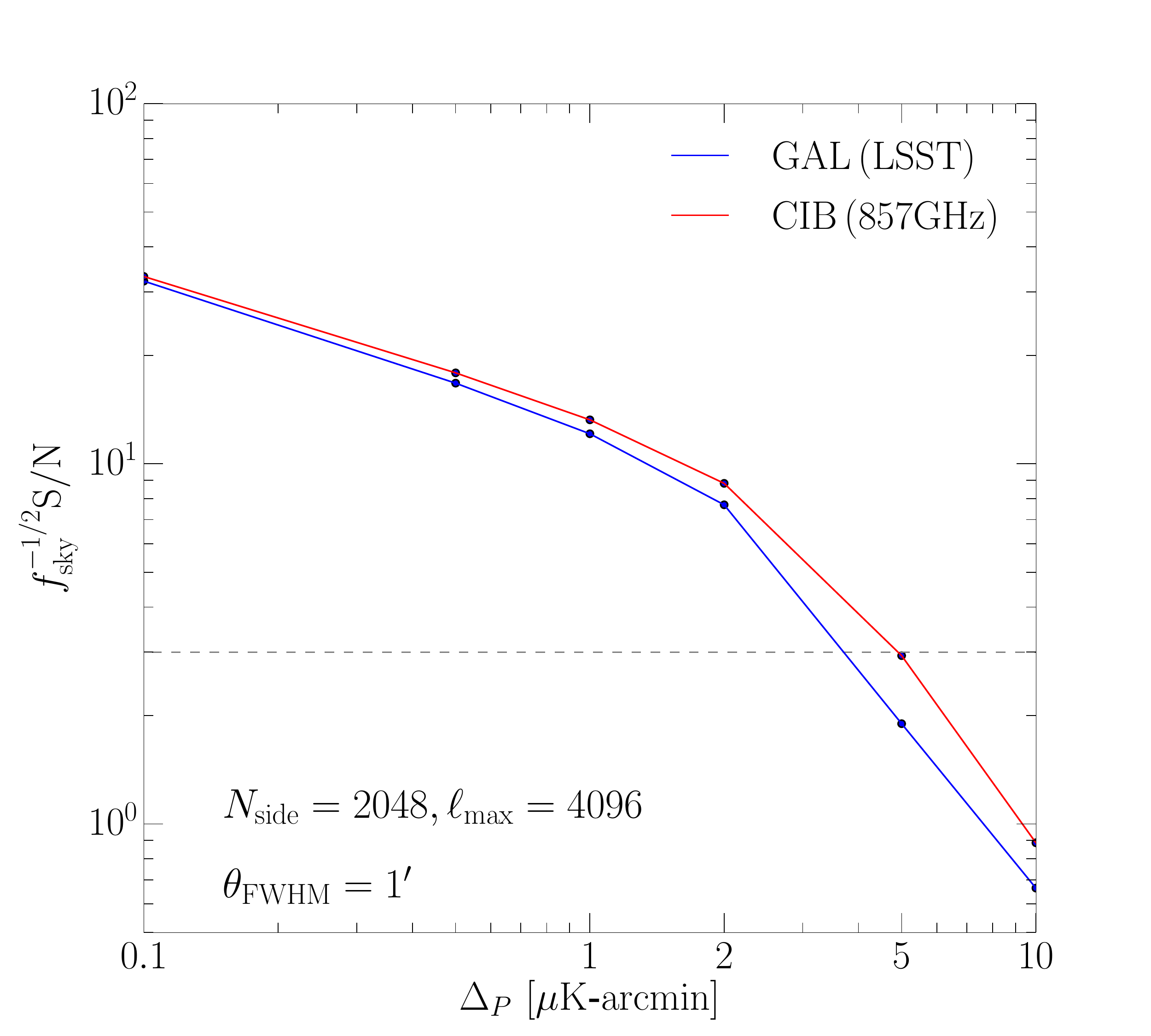}
\includegraphics[width=8cm, height=7.2cm]{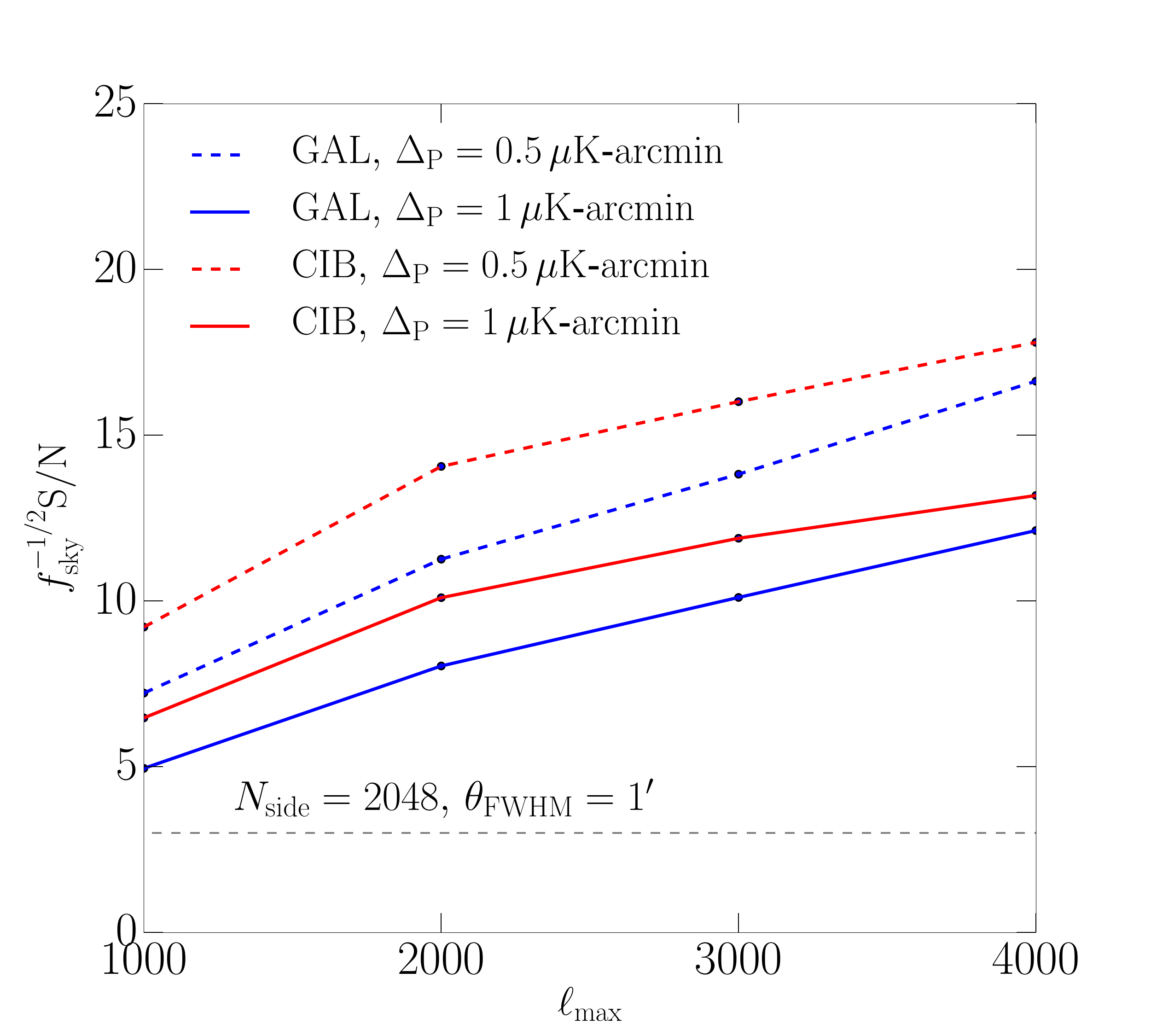}
\caption{Signal-to-noise ratios of $\langle\tau\Psi\rangle$ with respect to instrumental noise $\Delta_P$ (left) and $\ell_{\rm max}$ (right). 
The horizontal gray dashed line denotes a 3$\sigma$ detection threshold.}\label{sn2}
\end{figure*}

We forecast the detection significance for the $\langle\tau\Psi\rangle$ power spectrum.  For CMB-S4, we assume the polarization noise $\Delta_P$ = $\sqrt{2}\,\mu\rm{K}\mbox{-}\rm{arcmin}$, the full-width-half-maximum (FWHM) $\theta=1'$, and the sky fraction is $f_{\rm sky}$ = 0.5. The LSST galaxy survey will cover 18000 $\rm{deg}^2$, i.e., $f_{\rm sky}$ = 0.44. We use the redshift distribution in Ref.~\cite{2017arXiv171009465S} and derive the mean galaxy number per $\rm{arcmin}^2$ which is $\bar n$ = 67 $\rm{arcmin}^{-2}$. We simulate all maps with \textit{Healpix} at $N_{\rm side}$ = 2048 and $\ell_{\rm max}=4096$ and apply a 44\% mask to the simulations. The $\langle EBg\rangle$ and $\langle EB\Theta\rangle$ can both make $\sim 7\sigma$ and $\sim8\sigma$ detection of the cross-power spectra $\langle\tau g\rangle$ and $\langle\tau\Theta\rangle$, respectively.

\begin{table}
\caption{The $\phi$-type bispectrum $\textbf{b}^{(\phi)}(E,B,\Psi)$. The noise terms of CMB polarization and tracer are included in the power spectra in the denominators.}
  \begin{tabular}{ c|l | c | r }
    \hline
    $i$&$\alpha_{\ell}^{(\phi,i)}$&$\beta_{\ell}^{(\phi,i)}$&$\gamma_{\ell}^{(\phi,i)}$\\
    \hline
    0&$\frac{\ell(\ell+1)\tilde C_{\ell}^{EE}}{2C_{\ell}}$ & $\frac{1}{C_{\ell}}$ & $\frac{C_{\ell}^{\phi\Psi}}{C^{\Psi\Psi}_{\ell}+N_{\ell}^{\Psi\Psi}}$ \\ \hline
    1&$\frac{\tilde C_{\ell}^{EE}}{C_{\ell}}$ & -$\frac{\ell(\ell+1)}{2C_{\ell}}$ & $\frac{C_{\ell}^{\phi\Psi}}{C^{\Psi\Psi}_{\ell}+N_{\ell}^{\Psi\Psi}}$ \\ \hline
        2&$\frac{\tilde C_{\ell}^{EE}}{C_{\ell}}$ & $\frac{1}{C_{\ell}}$ & $\frac{\ell(\ell+1)C_{\ell}^{\phi\Psi}}{2(C^{\Psi\Psi}_{\ell}+N_{\ell}^{\Psi\Psi})}$ \\ \hline
  \end{tabular}\label{phiw}
\end{table}

\begin{table}
\caption{The $\tau$-type bispectrum $\textbf{b}^{(\tau)}( E,B,\Psi)$. The noise terms of CMB polarization and tracer are included in the power spectra in the denominators.}
  \begin{tabular}{ c|l | c | r }
    \hline
    $i$&$\alpha_{\ell}^{(\tau,i)}$&$\beta_{\ell}^{(\tau,i)}$&$\gamma_{\ell}^{(\tau,i)}$\\
    \hline
    0&$\frac{\tilde C_{\ell}^{EE}}{C_{\ell}}$ & $\frac{1}{C_{\ell}}$ & $\frac{C_{\ell}^{\tau\Psi}}{C^{\Psi\Psi}_{\ell}+N_{\ell}^{\Psi\Psi}}$ \\ \hline
  \end{tabular}\label{tauw}
\end{table}

\begin{table}
\caption{The experimental specifications}
  \begin{tabular}{ ccccc}
    \hline
  & CMB-S4&AdvACT&SPT-3G&SA (wide)\\
    \hline
    $f_{\rm sky}$&0.5&0.5&0.06&0.4\\
    \hline
   $\Delta_P\,[\mu{\rm K}\mbox{-}{\rm arcmin}]$&$\sqrt{2}$&10&2.5&5.5\\
   \hline
   $\theta_{\rm FWHM}\,[{}^{\prime}]$&1&1.4&1.2&3.5\\
    \hline
    SNR (GAL)&7$\sigma$&0.5$\sigma$&1.3$\sigma$&1.0$\sigma$\\
    \hline
    SNR (CIB)&8$\sigma$&0.6$\sigma$&1.7$\sigma$&1.6$\sigma$\\
        \hline
  \end{tabular}\label{snexp}
\end{table}

In Figure \ref{sn2} (left), we investigate the relationship between instrumental noise in polarization data and the overall the signal-to-noise ratio (SNR) of the $\langle\tau\Psi\rangle$ cross-power spectrum. We use the experimental specifications of CMB-S4, AdvACT~\cite{2016JLTP..184..772H}, SPT-3G~\cite{2014SPIE.9153E..1PB} and Simons-Array (SA) Wide~\cite{2016JLTP..184..805S} to forecast the detection significance at $N_{\rm side}$ = 2048, $\ell_{\rm max}=4096$ and $1^{\prime}$ beam. All the SNRs are given in Table \ref{snexp}. We find that a $3\sigma$ detection can be achieved when $\Delta_P< 4\,\mu\rm{K}\mbox{-}\rm{arcmin}$. In the right panel of Figure \ref{sn2}, we study the relationship between $\ell_{\rm max}$ and the SNR which can be further increased when more modes are used for both tracers.  
Moreover, we use numerical simulations to check the $f_{\rm sky}$ impact on the $\langle\tau \Psi\rangle$ detection significance which is found to be proportional to $f_{\rm sky}^{1/2}$. 

The successful reconstructions of $\langle\tau\Psi\rangle$ with both galaxy and CIB tracers verify that the bispectrum estimator for \pr\ and LSS cross correlations is applicable to a broad range of tracers, which can effectively enhance the detectability of the signatures of the \pr\ in the secondary CMB fluctuations. 
Given the fact that a spatially varying optical depth contains fluctuations from matter density and ionizing field and redshift distributions of the LSS tracers are limited to finite ranges, cross correlation $\langle\tau \Psi\rangle$ can precisely constrain matter-density-induced optical depth fluctuations but probably hard to probe the ionizing-field-induced ones which peak at very high redshifts. However, once a high redshift tracer is obtained, both contributions will be precisely captured by the cross correlation $\langle\tau \Psi\rangle$.

\section{Conclusion}

In this \pp\, we establish a bispectrum formalism to reconstruct the secondary signatures generated by \pr. The unbiased cross correlations between \pr\ and LSS tracers can be detected from next generation CMB experiments and LSS surveys. We further study the detection significance of the \pr\ from a few future CMB experiments in conjunction with tracers of large scale structure, such as LSST and CIB, and find that a $3\sigma$ detection can be achieved  
for high-resolution CMB polarization measurements with noise levels $\Delta_P< 4\,\mu\rm{K}\mbox{-}\rm{arcmin}$ and wide-area galaxy surveys. Furthermore, this method can be even extended to incorporate multiple tracers with redshift information so the detectability of the \prss\ signal will be dramatically improved, making the cross correlation $\langle\tau \Psi\rangle$ a new probe of cosmology and astrophysics.

\section{Acknowledgments}
We thank Brian Keating and Asantha Cooray for helpful discussions. 
This research 
is supported by the Brand and Monica Fortner Chair, and is part of the Blue Waters sustained-petascale computing project, which is supported by the National Science Foundation (awards OCI-0725070 and ACI-1238993) and the state of Illinois. Blue Waters is a joint effort of the University of Illinois at Urbana-Champaign and its National Center for Supercomputing Applications. We also acknowledge the use of the \hp~\cite{hp} package.

\bibliography{ebg}

\end{document}